\documentclass[10pt]{article}

\usepackage[OE]{express}
\usepackage{amsmath,amssymb}
\usepackage{tikz}
\usepackage{bm}
\usepackage{geometry}
\usepackage{epstopdf}
\usepackage{here}

\makeatletter
\g@addto@macro\normalsize{%
	\setlength\abovedisplayskip{0pt}
	\setlength\belowdisplayskip{10pt}
	\setlength\abovedisplayshortskip{0pt}
	\setlength\belowdisplayshortskip{10pt}
}
\makeatother

\begin{document}
	
	\title{Computational polarimetric microwave imaging}
	
	\author{Thomas Fromenteze,\authormark{1,2,*}  Okan Yurduseven,\authormark{2} Michael Boyarsky,\authormark{2}  Jonah Gollub,\authormark{2}  Daniel L. Marks,\authormark{2} and David R. Smith\authormark{2}}
	
	\address{\authormark{1}Xlim Research Institute, University of Limoges, 87060 Limoges, France\\
		\authormark{2}Center for Metamaterials and Integrated Plasmonics, Department of Electrical and Computer Engineering, Duke University, Durham, 27708, North Carolina, USA}
	
	\email{\authormark{*}thomas.fromenteze@xlim.fr}
	
	\newcommand{\bb}[1]{\bar{#1}}
	\newcommand{\subtr}[0]{_{\mathrm{t},\mathrm{r}}}
	
	\begin{abstract}
		We propose a polarimetric microwave imaging technique that exploits recent advances in computational imaging. We utilize a frequency-diverse cavity-backed metasurface, allowing us to demonstrate high-resolution polarimetric imaging using a single transceiver and frequency sweep over the operational microwave bandwidth. The frequency-diverse metasurface imager greatly simplifies the system architecture compared with active arrays and other conventional microwave imaging approaches. We further develop the theoretical framework for computational polarimetric imaging and validate the approach experimentally using a multi-modal leaky cavity. The scalar approximation for the interaction between the radiated waves and the target---often applied in microwave computational imaging schemes---is thus extended to retrieve the susceptibility tensors, and hence providing additional information about the targets. Computational polarimetry has relevance for existing systems in the field that extract polarimetric imagery, and particular for ground observation. A growing number of short-range microwave imaging applications can also notably benefit from computational polarimetry, particularly for imaging objects that are difficult to reconstruct when assuming scalar estimations.
	\end{abstract}
	
	
	\bibliographystyle{osajnl}
	\bibliography{biblio}
	
	\section{Introduction}
	Recent advances by the microwave community have resulted in the development of innovative imaging modalities for medical diagnosis~\cite{fear2002confocal,rubaek2007nonlinear,williams2008breast,nikolova2011microwave,bellizzi2011microwave,abbosh2016differential}, concealed threat detection~\cite{sheen2001three,sheen2010near,zhuge2011sparse,ahmed2011novel,ahmed2012advanced,rodriguez2014use,gonzalez2016millimeter}, through-wall imaging~\cite{baranoski2008through,dehmollaian2009through,ralston2010real,wang2012advanced}, and non-destructive testing~\cite{caorsi2004improved,benedetti2006innovative,kharkovsky2007microwave,hasar2009non}. Radio-frequency~(RF) bands are particularly suited to these applications, since RF waves can penetrate through many materials that are opaque at optical frequencies. Furthermore, since RF waves are non-ionizing, they are considered safe for human exposure at suitably low power levels. 
	
	The potential for efficient, cost-effective, and high-resolution systems that can achieve fast acquisition rates have recently been demonstrated in computational imaging systems based on cavity-backed ~\cite{carsenat2012uwb,fromenteze2015computational,fromenteze2016single} and metasurface ~\cite{hunt2013metamaterial,lipworth2013metamaterial,yurduseven2016frequency,marks2016spatially,gollub2017large,yurduseven2017millimeter} apertures. These systems radiate pseudo-orthogonal field distributions in transmission and---by exploitation of the reciprocity principle---in reception, to multiplex information and reconstruct an image. The frequency-diverse aperture limits the complexity of the hardware architecture required for real-time high-resolution imaging, obviating the actively controlled components or the need for mechanical motion that is typically required in conventional systems. Recently, frequency-diverse metasurface apertures were adapted for intensity-only measurements, demonstrating that the phase information can be coded in the dispersive radiation of these complex, but passive, antennas and solved for within the inverse problem ~\cite{fromenteze2016phaseless}. This allows further simplification of the associated measurement electronics. However, the application of such computational systems to polarimetric imaging remains unexplored. 
	
	Polarization is a relevant source of information in radar imaging, allowing determination of the nature of the interaction between in scene targets and the spatial components of the radiated electromagnetic fields ~\cite{chew1995waves}. Polarimetric radars were initially and widely implemented in geoscience applications, allowing for the interrogation of relevant parameters unavailable to single-polarized radars~(soil classification, soil moisture, snow distribution, etc.)~\cite{ulaby1990radar,treuhaft2000vertical,lee2009polarimetric,velotto2013dual}. \textcolor{black}{Following the growing interest in short-range imagery, polarization sensitive radars have been adapted for exploiting the vectorial nature of electromagnetic waves. These approaches thus improve the quality of the reconstructed images, enhancing the accuracy of the estimated contrasts and revealing target features invisible to the approaches based on single-polarized measurements~\cite{amineh2012three,salman2012polarization,wang2016novel}. The elimination of artifacts due to multipath in concealed weapon detection has also been demonstrated by illuminating targets using a combinations of left and right-hand circular polarizations~\cite{sheen2005circularly,sheen2005circularlyb}. Finally, full-polarimetric imaging has found recent applications in medical diagnosis~\cite{abubakar2002imaging}.} However, we note that the technical constraints associated with the development of polarimetric radars are exacerbated when striving for high-resolution and real-time systems, therefore making full reconstruction of polarimetric images extremely burdensome.
	
	In this paper, we propose to extend the framework of microwave computational imaging to the measurement of polarization information in short-range applications by encoding the susceptibility of the target in the physical layer of the antenna, i.e. as a single frequency-dependent signal. In the \textcolor{black}{Section 2} of this article, the theoretical principle of polarimetric computational imaging is first introduced. A generic polarization-sensitive metasurface is considered for simplifying the architecture required for retrieving vectorial information in the target space. \textcolor{black}{The cavity-backed metasurface developed to this end is then presented in the Section 3}. A proof of concept of short-range imaging is then presented for validating this novel principle \textcolor{black}{in the Section 4}. \textcolor{black}{Finally Section 5 provides the concluding remarks.}
	
	\section{Computational polarimetry: the interrogation of a susceptibility tensor from a signal coded in the physical layer}
	
\begin{figure}[ht]
		\centering
		\includegraphics[width = 0.8\textwidth]{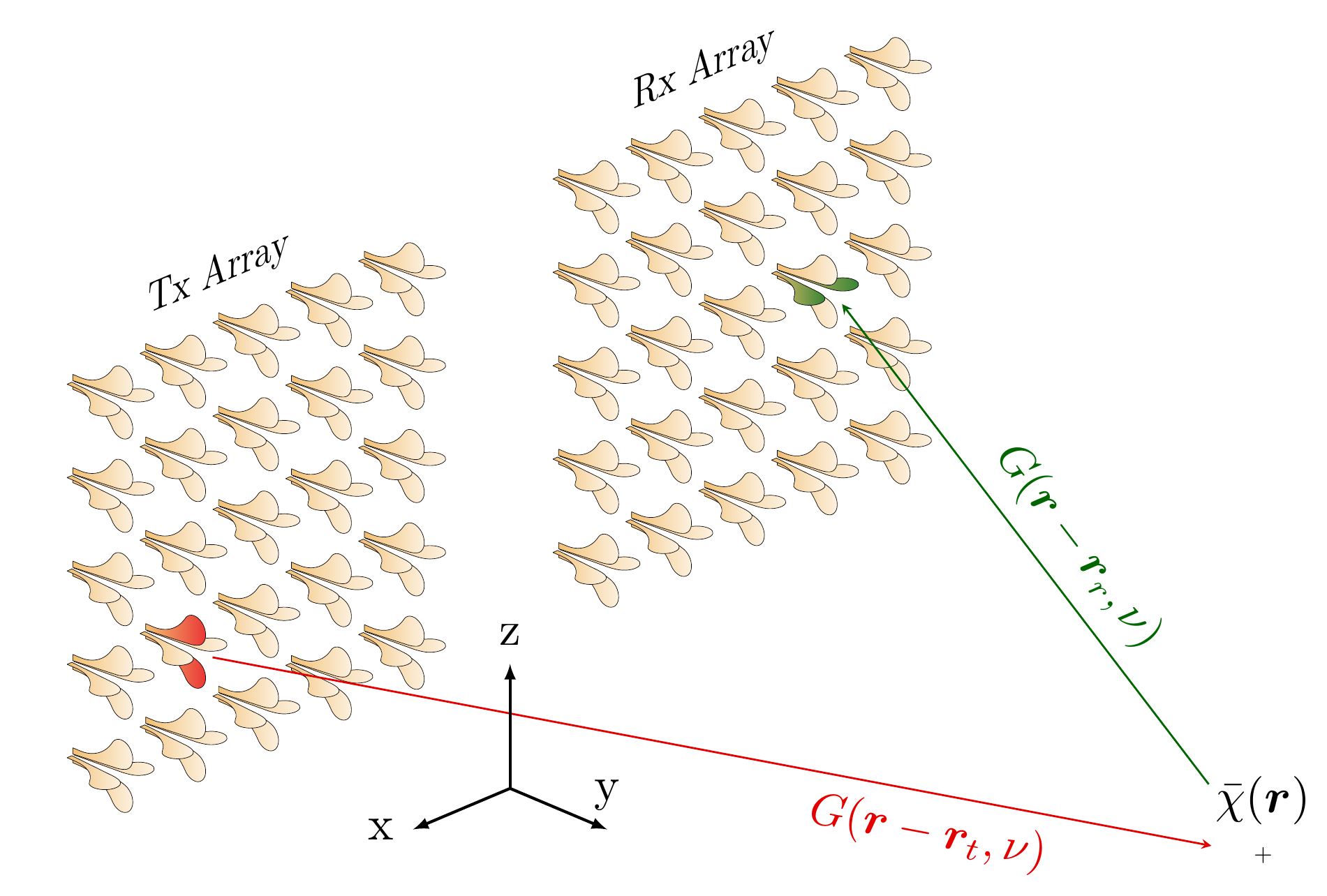}
		\caption{Polarimetric radar system: Two arrays made of dual-polarized antennas are used in transmission and reception for interrogating the susceptibility tensor $\bar \chi(\bm r)$ in the target space. To estimate the image, the fields radiated by each antennas must be known---either measured or derived (e.g. from the propagation of the surface currents with the dyadic Green's functions $G(\bm r- \bm r\subtr\, ,\nu)$.} 
		\label{fig:PolarArray}
	\end{figure}

	We first describe the underlying principle of polarimetric microwave imaging, using the illustrative geometry shown in Fig. ~\ref{fig:PolarArray}. 
	
	In this conceptual example, two arrays of dual-polarized antennas are used to measure the interaction between radiated waves and a target. In the configuration shown, the field is generated by the transmit array and the scattered fields are detected by the receive~array. The electric susceptibility tensor $\bb \chi( r)$ accounts for the interaction between the radiated waves and the illuminated scene. In a conventional computational imaging approach, the field radiated and measured by a pair of transmit and receive antennas would interact with the target susceptibility, generating the frequency-dependent signal $\bar S(\bm r_t, \bm r_r, \nu)$. This $2 \times 2$ matrix contains the interaction between the excited $x$ or $z$-oriented element in transmission at location $\bm r_t$ with the $x$ or $z$-oriented element in reception at location $\bm r_r$, where $x$ and $z$ are coordinates in the plane of the arrays (both arrays are assumed to be in the same plane). In the first Born approximation, the interaction between the dyadic fields radiated by the $x$-polarized antenna in transmission, with the $z$-polarized antenna in reception is: 
	
	\begin{align}
	S_{x,z}(\bm r_t, \bm r_r, \nu) = \int_r \bm E_x (\bm r_t,\bm r, \nu)\, \bb \chi (\bm r)\,\bm E_z (\bm r_r,\bm r, \nu)^T\ d^3 r
	\end{align}
	
	The vector fields $\bm E_x (\bm r_t, r, \nu)$ and  $\bm E_z (\bm r_r, r, \nu)$ are radiated by each of the transmit and receive antennas from their respective locations $\bm r_t$ and $\bm r_r$. The total measured frequency-dependent signal can be expressed in matrix form for each dual-polarized transmit and receive pair as:
	
	\begin{align}
	\bb S(\bm r_t, \bm r_r, \nu) = \int_r \bar E (\bm r_t,\bm r, \nu)\, \bb \chi (\bm r)\,\bar E (\bm r_r,\bm r, \nu)^T\ d^3 r
	\label{eq:DualPolar}
	\end{align}
	
	\noindent where $\bar E$ is a $2 \times 3$ matrix representing the three-dimensional vector field radiated by two co-localized transverse polarized antennas.
	
	Assuming that these fields are determined either by an experimental measurement or from a numerical analysis, the susceptibility tensor can then be interrogated by solving the inverse problem corresponding to the forward model specified by~Eq.~(\ref{eq:DualPolar}). Controlling spatially the radiated polarized fields over two apertures is particularly burdensome in applications where real-time and high-resolution constraints are imposed. The RF architectures associated with each array would thus have to include as many active chains as dual-polarized antennas, working simultaneously over an ultra-wide bandwidth. Fast switches would also need to be implemented for selecting the excited polarization of each antenna leading to three-dimensional vector fields.
	
\begin{figure}[ht]
		\centering
		\includegraphics[width=0.9\linewidth]{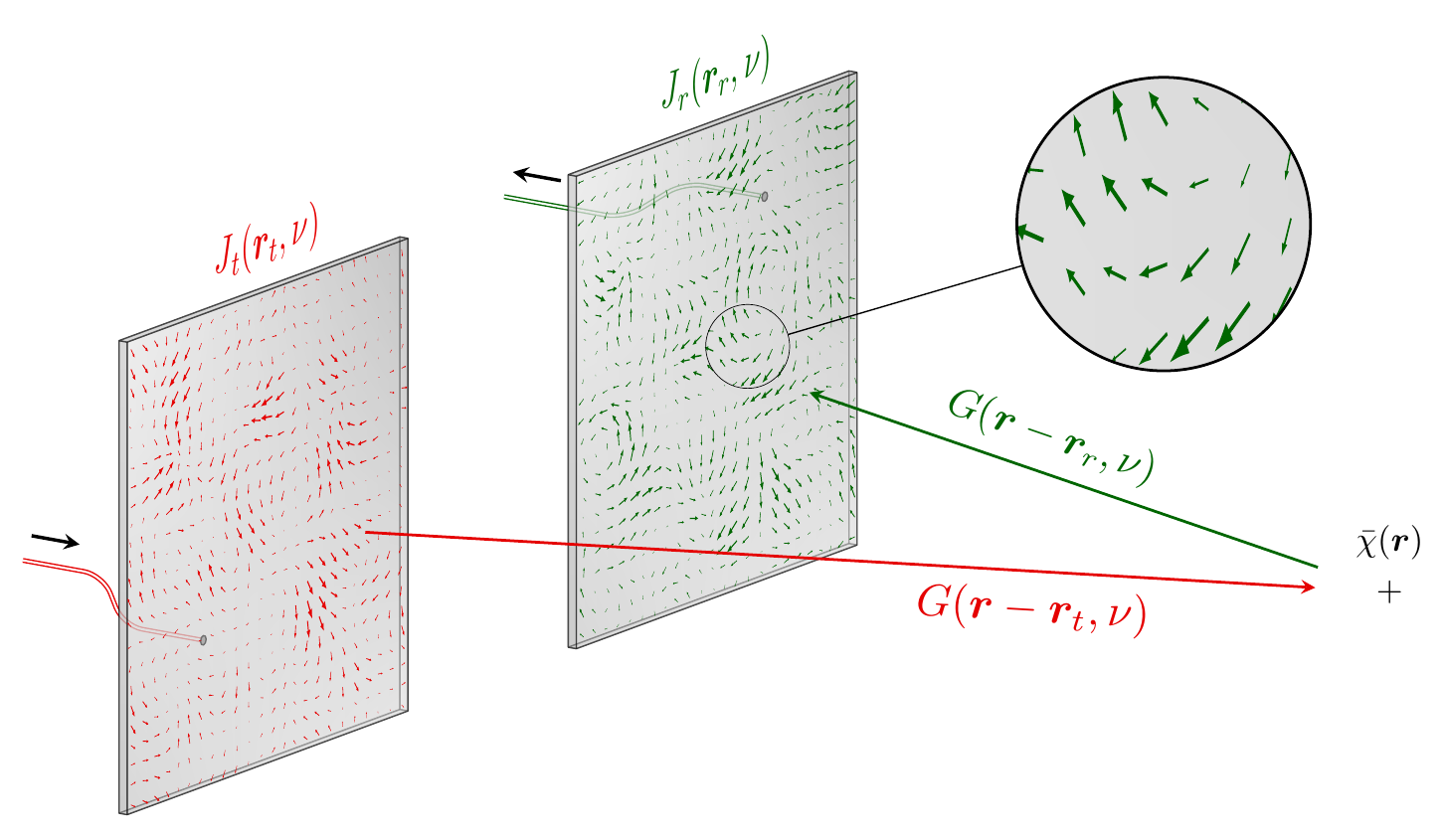}
		\caption{Radiating metasurface implemented for the demonstration of microwave computational polarimetry. The current distribution is shown on the aperture and represents an excitation of the transmission port at a single frequency. The metasurface is conceived to obtain pseudo-orthogonal current distributions in the frequency domain.}
		\label{fig:BothPanels}
	\end{figure}

	In this paper, the polarimetric imaging approach is simplified by implementing a computational technique (Fig.~\ref{fig:BothPanels}).
	
	The two arrays are replaced by metasurfaces, which radiate pseudo-orthogonal patterns in space and frequency. \textcolor{black}{ In this way, the large number of active chains implemented for the independent transmission and reception of waves (multiplied by two for a polarization sensitive system) is substantially decreased. Instead a single compressed signal can be measured through the input and output ports and the radiating antennas. The amount of information included in a set of these measurements directly depends on the spatial diversity of the fields emitted and received by the antennas~\cite{marks2016spatially}, and is now studied considering the potential contribution of polarization diversity.}
	
	The expression of the fields $\bm E(\bm r\subtr , \bm r,\nu)$ radiated respectively by the transmit and receive arrays is defined by the expression of the equivalent vector potentials $\bm A\subtr (\bm r\subtr, \nu)$, according to the corresponding current distributions $\bm J\subtr(\bm r\subtr, \nu)$ in the radiating apertures that $\bm A\subtr (\bm r\subtr, \nu)$ satisfies the Helmoltz equation~\cite{chew1995waves}:
	
	\begin{align}
	\nabla^2 \bm A\subtr(\bm r\subtr,\nu)  + k^2 \bm A\subtr(\bm r\subtr,\nu)  = - \mu_0\ \bm J\subtr(\bm r\subtr,\nu) 
	\label{eq:PotentialVector}
	\end{align}
	
	\noindent The expression of $\bm A\subtr (\bm r,\nu)$ is thus obtained by computing the convolution product between the current distribution \bm $\bm J\subtr(\bm r\subtr,\nu)$ and a scalar free-space Green's function $g(\bm r,\nu)$ in the antenna space $\bm r\subtr$:
	
	\begin{align}
	\bm A\subtr(\bm r,\nu) = \mu_0 \int_{\bm r\subtr} \bm J\subtr(\bm r\subtr,\nu)\ g(\bm r - \bm r\subtr,\nu)\ d^3\bm r\subtr
	\end{align}
	
	\noindent The expression of the electric field is finally given by~\cite{chew1995waves}:
	
	\begin{align}
	\label{eq:Efield}
	\bm E\subtr(\bm r,\nu) &= -\frac{j c}{k} \left[ \nabla(\nabla \cdot \bm A\subtr(\bm r,\nu)) + k^2 \bm A\subtr(\bm r,\nu) \right]\\
	&= j 2 \pi \nu \mu_0 \int_{\bm r\subtr} \bm J\subtr(\bm r\subtr,\nu)\ \bb{G}(\bm r, \bm r\subtr,\nu)\  d^3\bm r\subtr
	\end{align}
	
	\noindent where $k = 2 \pi \nu/c$ is the wavenumber and $\bb{G}(\bm r_a, \bm r_b,\nu)$ is the dyadic Green's function modeling the propagation from $\bm r_a$ to $\bm r_b$, written as:
	
	\begin{align}
	\bb{G}(\bm r_a, \bm r_b,\nu) = \left(\bb{I} + \frac{\nabla\nabla}{k^2} \right)\{g(\bm r_a - \bm r_b,\nu)\}
	\end{align}
	
	%
	
	The susceptibility tensor $\bb\chi(\bm r)$ can be estimated from the interaction between the fields radiated by two metasurfaces~\cite{marks2016spatially}, leading to the following measured frequency signal: 
	
	\begin{align}
	\label{eq:MeasuredSignal}
	s(\nu) &= \frac{j \pi \epsilon_0}{\nu}\int_r \bm E_\mathrm{r}(\bm r,\nu)\, \bb\chi(\bm r)\, \bm E_\mathrm{t}(\bm r,\nu)^T\, d^3\bm r\\
	&= \frac{j \pi \epsilon_0}{\nu}\int_r \mathrm{Tr}[ \bb\chi(\bm r)\, \bm E_\mathrm{r}(\bm r,\nu)\, \bm E_\mathrm{t}(\bm r,\nu)^T]\ d^3\bm r
	\end{align}
	
\begin{figure}[ht]
		\centering
		\includegraphics[width=0.49\textwidth]{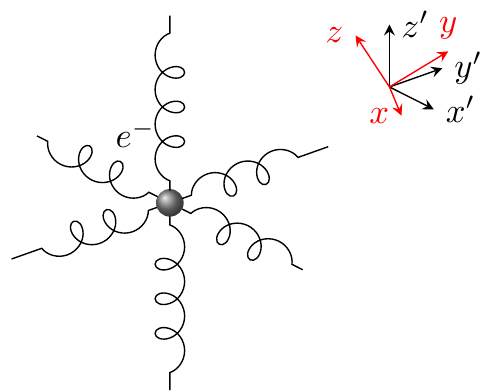}
		\caption{Description of the susceptibility tensor with a mechanical model of the bound electron. The induced polarization corresponds to a motion of these particles reacting to an electric field. This tensor is diagonalized by the compensation of the rotation between the radiated wave's axis (red) and the main axis of the electron (black) confined in its movements by the interaction with the surrounding particles.}
		\label{fig:SpringModel}
	\end{figure}

	\noindent The symmetry of the dyadic Green's function matrix leads to $\bb{G}(\bm r, \bm r_\mathrm{t},\nu) = \bb{G}(\bm r, \bm r_\mathrm{t},\nu)^T$. As demonstrated in~\cite{marks2016spatially}, the outer product $\bm E_\mathrm{r}(\bm r,\nu)\, \bm E_\mathrm{t}(\bm r,\nu)^T$ forms a basis interrogating the reflectivity of the scene. In previous studies of computational imaging using a signal coded by the physical layer, a scalar approximation was used for the susceptibility tensor, allowing for an estimation of a target reflectivity illuminated by several metasurfaces. Here, we extend this framework to a tensorial reconstruction of the susceptibility in the target space, passively encoding the three-dimensional polarization information into a unique frequency signal measured between two ports. This signal is simplified when restricting the study to anisotropic media defined by a diagonalized susceptibility tensor.	From a fundamental perspective, we consider electrons within the target medium interacting with the incident electric field as being represented by a simple mechanical model (Fig.~\ref{fig:SpringModel}).
	
\begin{figure}[ht]
		\centering
		\includegraphics[width = 0.75\textwidth]{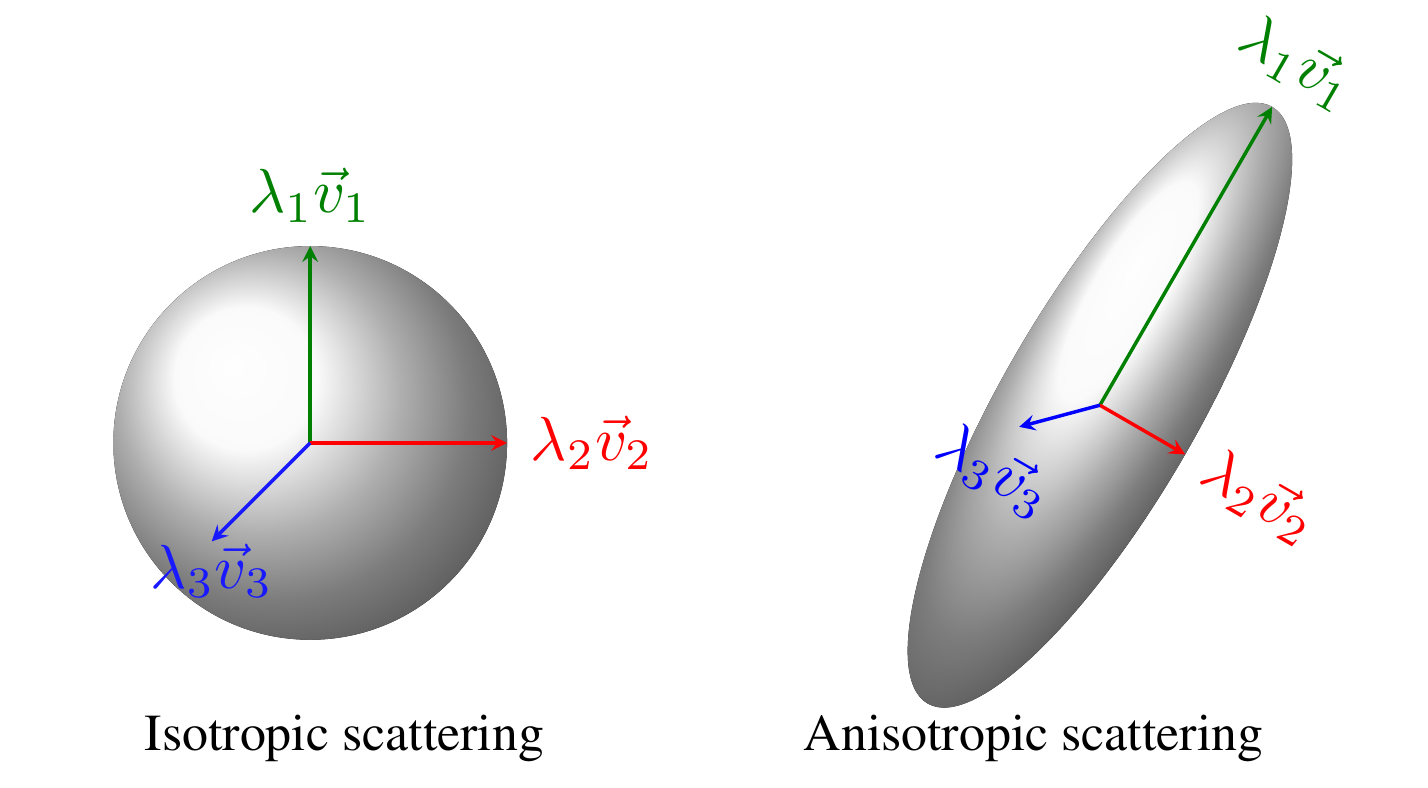}
		\caption{Ellipsoids representing the eigenvalues/eigenvectors decomposition of the electric susceptibility tensor estimated for each voxel. The anisotropic scattering is represented by an ellipsoid with axis of unequal lengths, depicting the local orientation of the tensor.}
		\label{fig:Ellipsoids}
	\end{figure}

	In the depicted model, the principal axes of the bound electron ($x',y',z'$) are not aligned with the coordinate axes ($x,y,z$), the latter defined as references for the transmitted and received electromagnetic fields. Thus, the non-diagonal terms are created by projections of the radiated field over the main axis of the tensor at each location, leading to the following eigendecomposition:
	
	\begin{align}
	\bb \chi(\bm r) = \bb R(\bm r)\  \text{diag}\big(\bm \xi(\bm r)\big)\ \bb R(\bm r)^T
	\end{align} 
	
	\noindent where $\bb R(\bm r)$ is a rotation matrix and $\xi(\bm r)$ is the diagonalized susceptibility tensor. Assuming the first Born approximation, a forward model is defined that links the measurements to the susceptibility tensor defined at each location of the target space. The principal axes of each tensor can thus be estimated from a diagonalization determining the orientation of the currents on the targets, which is homogenized over the size of the reconstructed voxels. \textcolor{black}{Finally, These estimations can be represented using the weighted eigenvectors or equivalent ellipsoids, showing the anisotropy of the scattering (Fig.~\ref{fig:Ellipsoids}). Similar studies are notably developed for the estimation of anisotropy in the the field of diffusion tensor imaging~\cite{basser1994mr,le2001diffusion,mori2006principles}}.
	
\begin{figure}[ht]
		\centering
		\includegraphics[width=\linewidth]{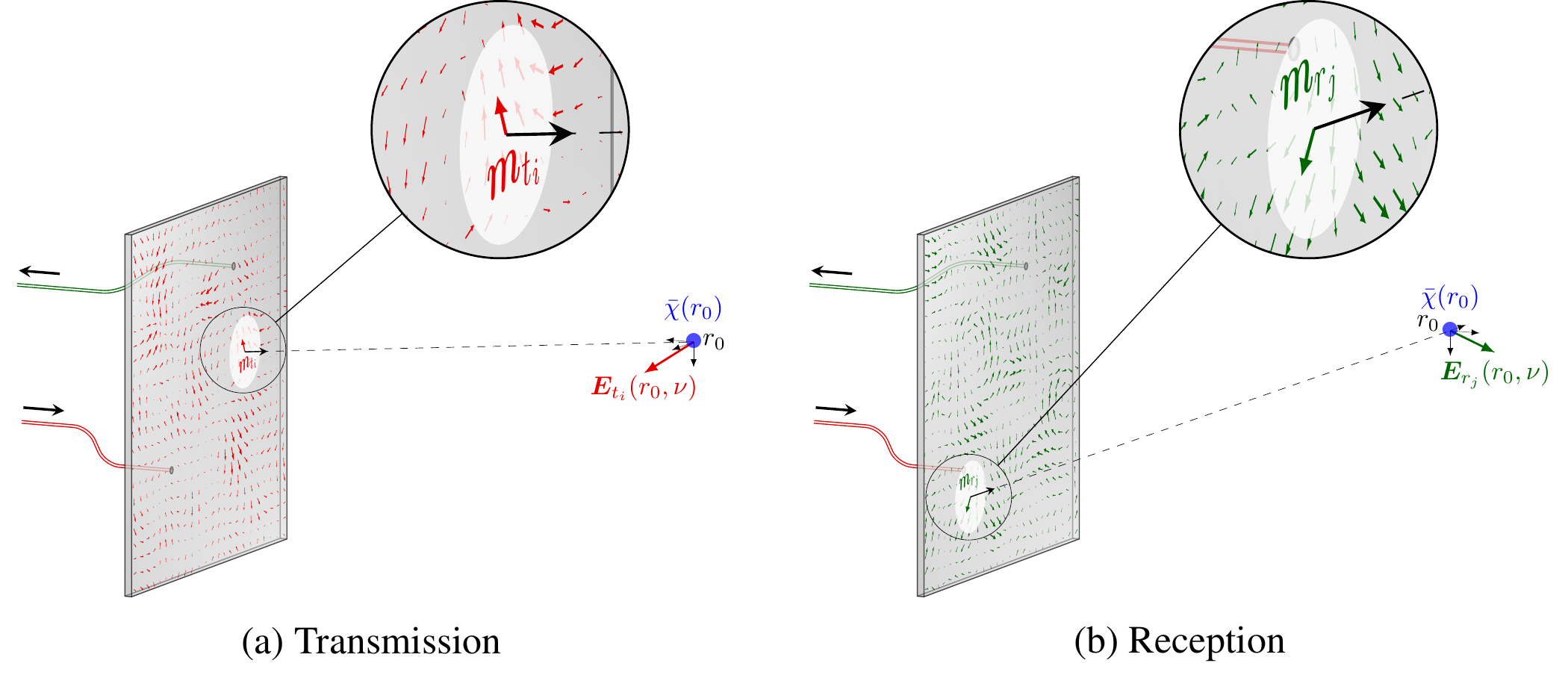}
		\caption{Analysis of the field generated by one dipole of the transmitting (a) and receiving (b) current distributions at the same location $\bm r_0$.}
		\label{fig:MetasurfacesPolar}
	\end{figure}

	Having introduced the essential aspects of the computational imaging, we take a closer look at the radiation of polarized fields by metasurfaces. Indeed, simplifications can be applied when considering antennas radiating in only one direction of the space. The radiated field can thus be deduced by the measurement of the tangential electric field on a synthetic aperture plane with vanishing boundary conditions, denoted $\bm E\subtr^{\mathrm{tan}}(\bm r, \nu)$~\cite{jackson1999classical}. From this field, we can extract a set of equivalent dipoles, as 
	
	\begin{align}
	\bm p\subtr (\bm r \subtr) &= \epsilon_0\,\bm n\int_{r^{'}\subtr} \left(\hat{\bm y} \cdot \bm E\subtr^{\mathrm{tan}} (\bm r^{'}\subtr, \nu) \right)\ d^2 \bm r^{'}\subtr\\
	\bm m\subtr (\bm r \subtr) &= \frac{1}{j \mu \pi \nu} \int_{r^{'}\subtr} \left(\bm n \times \bm E\subtr^{\mathrm{tan}} (\bm r^{'}\subtr, \nu) \right) \ d^2 \bm r^{'}\subtr
	\label{eq:Dipoles}
	\end{align}
	
	\noindent where $\bm p\subtr(\bm r \subtr) $ are the equivalent electric dipoles and $\bm m\subtr(\bm r \subtr) $ the magnetic ones, computed in transmission and reception over small 2D domains $\bm r^{'}\subtr$, effectively defining a grid where each facet is centered at $\bm r \subtr$. The vector $\bm n$ is defined as the normal to the measurement plane and $\hat{\bm y}$ is a unit vector perpendicular to the plane (optical axis). For the sake of simplicity, the transmit and receive metasurfaces can be defined as co-located and oriented such that $\bm n = \hat{\bm y}$ for each radiating plane. The electric dipoles $\bm p\subtr$ are thus canceled out by the dot product $\hat{\bm y} \cdot \bm E\subtr^{\mathrm{tan}} = 0$, allowing for a decomposition of the field in the target space as the superposition of contributions radiated by magnetic dipoles only, as depicted in Fig.~\ref{fig:MetasurfacesPolar}.
	
	
	In this example, different distributions of random magnetic dipoles are radiating according to the excitation port and to the frequency. The expression of the radiated electric field in the target space is finally expressed from the contributions of all the magnetic dipoles, summing over the discrete dipole space $\bm r \subtr$~\cite{jackson1999classical,lipworth2015comprehensive}:
	
	\begin{align}
	\bm E \subtr (\bm r, \nu) = -\frac{\mu_0 \pi \nu^2}{c} \sum_{\bm r \subtr} [\bm r \times \bm m_{\mathrm{t},\mathrm{r}}(\bm r \subtr)]\ g(\bm r\subtr- \bm r, \nu)
	\label{eq:discreteFields}
	\end{align}
	
	\noindent with the cross-product given as:
	
	\begin{align}
	\bm r \times  \bm m\subtr(\bm r\subtr)
	=
	\begin{bmatrix}
	0 & -z & y \\
	z & 0 & -x \\
	-y & x & 0
	\end{bmatrix}
	\begin{bmatrix}
	m_x\\
	0\\
	m_z
	\end{bmatrix}
	=
	\begin{bmatrix}
	y\ m_z\\
	z\ m_x - x\ m_z\\
	x\ m_y - y\ m_x
	\end{bmatrix}
	\end{align}
	
	The projection property of the cross product allows for a random distribution of magnetic sources contained in a plane to be sufficient for radiating 3D polarized electric near fields that are able to interrogate the full susceptibility tensor. A pseudo-random distribution of sources in the flat metasurface aperture thus radiates a complex field able to probe the medium of interest, encoding the susceptibility variation of the target space into a unique frequency-dependent signal. This system is able to passively multiplex the target space information in transmission and reception, extending earlier approaches based on a scalar approximation of the susceptibility tensor in~\cite{fromenteze2014passive,fromenteze2016single}.
	
	The susceptibility of the target space can be estimated by a pseudo-inversion of the radiated fields in transmission and reception of the form
	
	\begin{align}
	\hat{\bar{\chi}} (\bm r) = \sum_\nu \frac{\nu}{j \pi \epsilon_0}\ \bm E_\mathrm{r}(\bm r,\nu)^+ s(\nu)\  \bm E_\mathrm{t}(\bm r,\nu)^{+^T}
	\label{eq:Inv}
	\end{align}
	
	\noindent where $E\subtr(\bm r,\nu)^+$ is the pseudo-inverse computed for each independent polarization state on the discrete representation of the radiated fields in transmission and reception~\cite{barrett1994templates,yurduseven2015resolution,fromenteze2015unification}, determined from Eqs.~(\ref{eq:Dipoles}) and (\ref{eq:discreteFields}) and assuming that the tangential field $\bm E\subtr^{\mathrm{tan}} (\bm r, \nu)$ can be determined from an analytical model (or measured).
	
	We stress that these estimations can only be achieved efficiently if these fields present a low level of correlation, spatially, and for each polarization state. Conveniently, we can exploit frequency diversity to encode the measured information onto the ports connected to the metasurface antenna.
	
	To be more specific, we propose to develop the expression of the sub-element $\hat{\bar{\chi}}_{(m,n)}$ of the retrieved susceptibility tensor at the location $r_0$:
	
	\begin{align}
	\hat{\bar{\chi}}_{(m,n)} (r_0) &= \sum_\nu \frac{\nu}{j \pi \epsilon_0}\ E_{\mathrm{r},m}(r_0,\nu)^+ s(\nu)\ E_{\mathrm{t},n}(r_0,\nu)^{+^T}\\
	&= \sum_{p,q} \bar{\chi}_{(p,q)} (r_0)\, R_{(m,n,p,q)}(r_0)  
	\end{align}
	
	where $R_{(m,n,p,q)}(r_0)$ is a function of the degree of correlation in space and polarization of the radiated fields, and is defined as: 
	
	\begin{align}
	R_{(m,n,p,q)}(r_0) = \sum_\nu \int_r  E_{\mathrm{r},m}(r_0,\nu)^+ E_{\mathrm{r},p}(\bm r,\nu)\, E_{\mathrm{t},q}(\bm r,\nu)^T\,  E_{\mathrm{t},n}(r_0,\nu)^{+^T} d^3\bm r 
	\end{align}
	
	An ideal estimation of the full susceptibility tensor at this specific location would thus impose to ensure that 
	
	\begin{align}
	R_{(m,n,p,q)}(r_0) = 
	\begin{cases}
	1, &\text{if}\ (m,n)=(p,q)\\
	0 &\text{if}\ (m,n)\ne(p,q)
	\end{cases}
	\end{align}
	
	In practice, the sum of non-zero cross-terms can thus introduce a speckle noise impacting the accuracy of the estimation. In general, the reconstructions will thus highly depend on the technique implemented for solving the inverse problem formulated in Eq. (\ref{eq:Inv}), using direct or iterative strategies of reconstruction. In this article, all the results were computed by implementing an iterative least-square algorithm.
	
	\section{Cavity-backed polarization-sensitive metasurface}
	
	We validate the computational polarimetric imaging technique using the frequency-diverse, chaotic leaky microwave cavity illustrated in Fig.~\ref{fig:Cavity}. 
	
	\def \picheight {5.2cm}
	
	\begin{figure}[ht]
		\centering
		\includegraphics[height=\picheight]{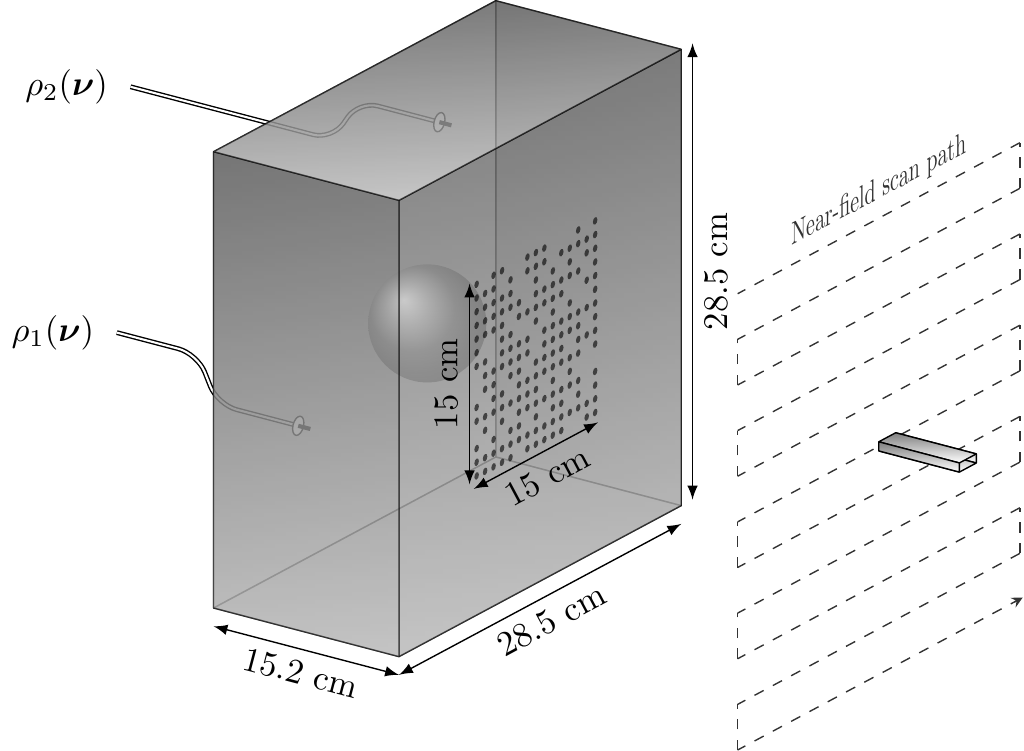}
		\hspace{0.2cm}
		\includegraphics[height=\picheight]{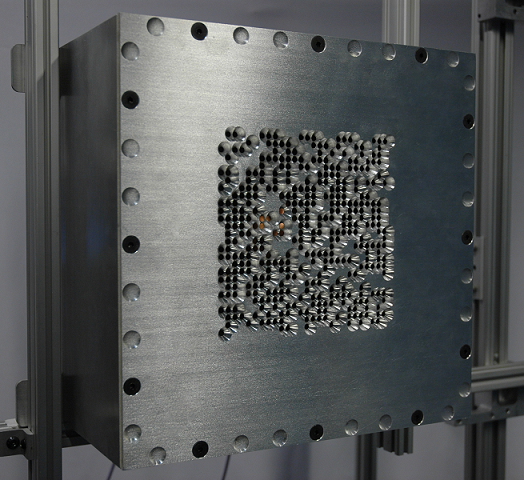}
		\caption{Radiating metasurface implemented for the validation of the proposed computational polarimetric imaging technique. \textcolor{black}{The radiated near-field is measured on both transverse polarizations with a single-polarized open-ended waveguide probe.}}
		\label{fig:Cavity}
	\end{figure}
	
\begin{figure}[ht]
		\includegraphics[width=1\textwidth]{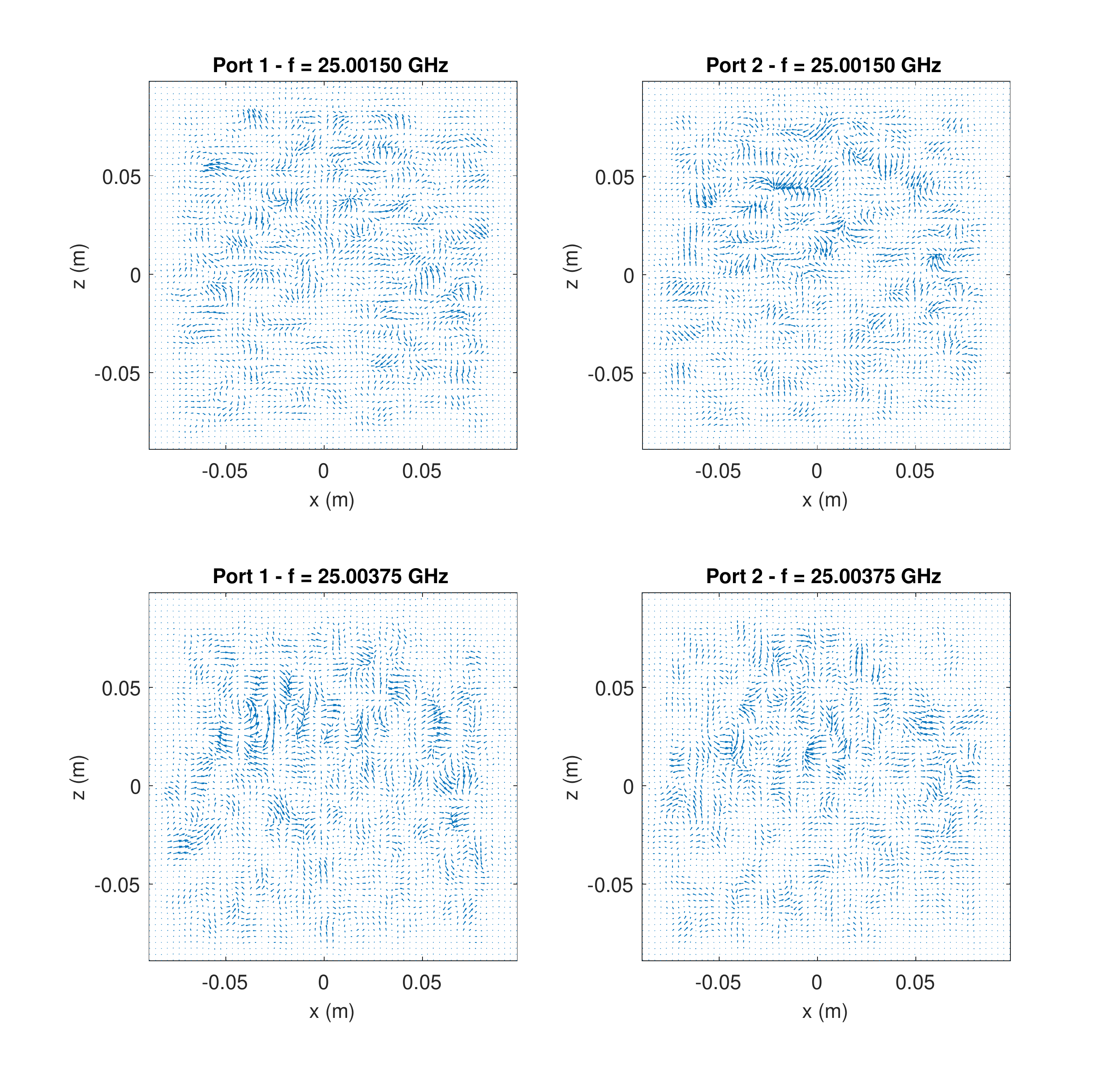}
		\vspace{-1cm}
		\caption{Near-field scans measured by a raster scan back-propagated to the aperture of the metasurface. The real part of the fields are represented for each excitation port and two consecutive frequencies just above 25 GHz, spaced by only 2.5 MHz.}
		\label{fig:ScansPol}
	\end{figure}

	The radiated patterns must exhibit a low level of correlation among the radiated field patterns in frequency, space, and polarization---ensuring an appropriate multiplexing of the polarimetric information gathered from the target space and encoding it into a single frequency vector. These properties are obtained by exploiting the modal diversity in a leaky cavity designed to be several times larger than the operating wavelength. The frequency bandwidth is thus set between $17.5$ and $26.5$ GHz, corresponding to wavelengths of 1.15 cm to 1.67 cm, for a cavity with inner dimensions of $15.2 \times 28.5 \times 28.5\ \text{cm}^3$. This type of over-sized reverberating cavity finds applications in various domains such as the evaluation of the gain of antennas or for electromagnetic vulnerability measurements~\cite{kildal2002definition,hill2002spatial}. A metallic ball is inserted into this air-filled cavity to optimize the modal diversity, to create irregular and convex boundaries and prevent the development of degenerate states~\cite{draeger1997one,montaldo2004time}. As opposed to previous microwave computational imaging demonstrations \cite{fromenteze2016phaseless,fromenteze2015computational} also based on air-filled leaky cavities, this work introduces a means of estimating the vector response of the target space instead of its scalar approximation. In addition to the low level of correlation of the fields in space and frequency that is required in the previous experiments, we also require an additional constraint of pseudo-orthogonality in polarization. The various correlations in frequency, space and polarization can be probed by analyzing the near field scans of $\bm E_{1,2}^{\mathrm{tan}} (\bm r\subtr, \nu)$ obtained by exciting the Ports 1 and 2 connected to the cavity successively (Fig.~\ref{fig:ScansPol}).
	
\begin{figure}[ht]
		\centering
		\includegraphics[width=0.8\textwidth]{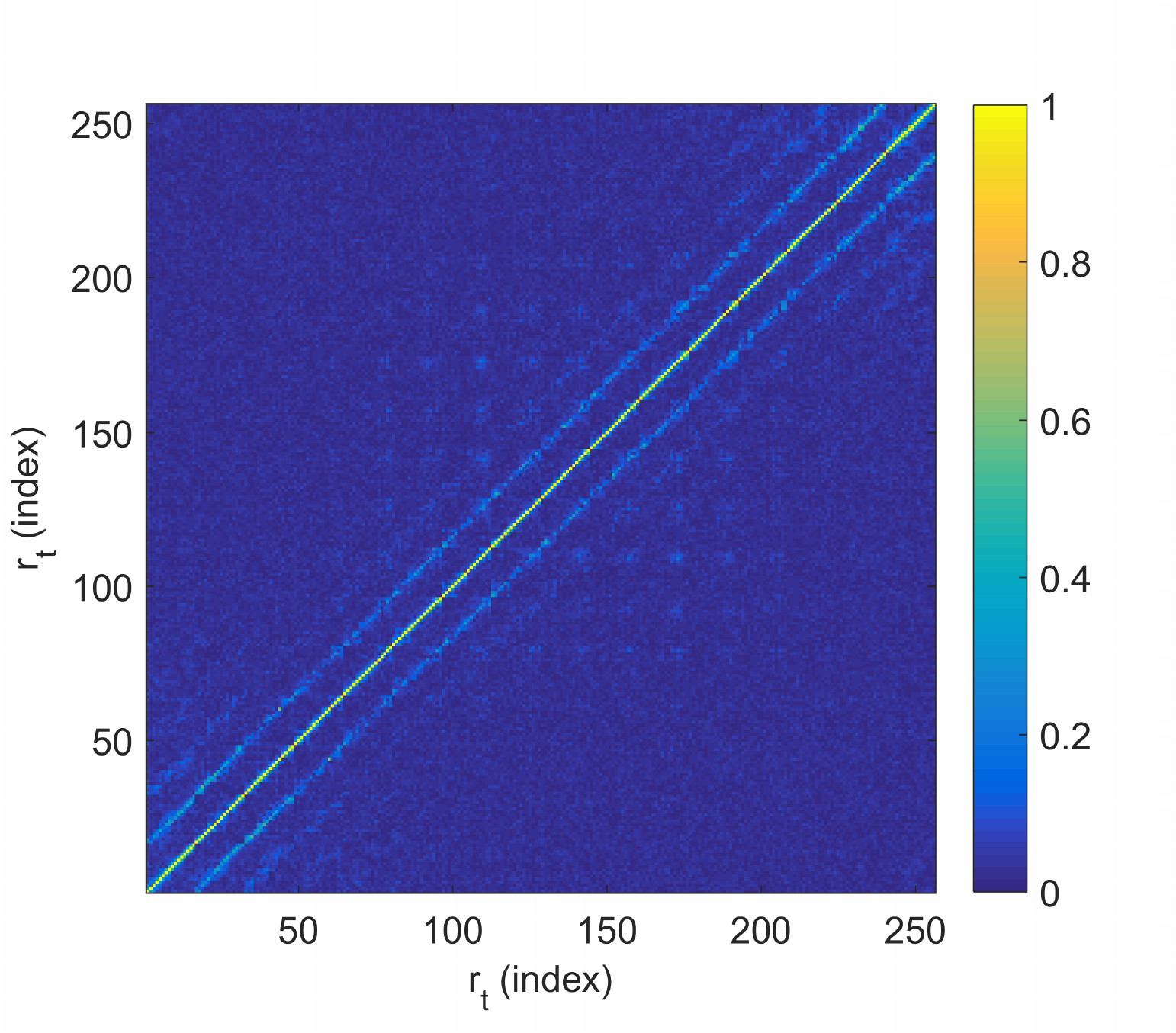}
		\caption{Autocorrelation of the processed near field scan obtained on the $x$-polarization when feeding the port 1 of the cavity.}
		\label{fig:Xcorr11}
	\end{figure}

	For the measurements, a commercial near-field scanner was used, NSI 200V-3x3 \cite{fromenteze2015computational,yurduseven2016frequency}. The measured field distribution is equivalent to the one depicted in Fig.~\ref{fig:MetasurfacesPolar}, since the tangential fields and the magnetic dipoles are directly related (Eq.~\ref{eq:Dipoles}). One can observe that both the location of the feeding port and the frequency, even when the shift represents a tiny fraction of the operating frequency, have a noticeable impact on the near field distribution in the aperture of the metasurface. Before the introduction of any imaging experiment, the correlation between the radiated fields is studied. This analysis is initiated by considering only the field radiated on the $x$-polarization when feeding port 1. Near field scans are recorded over 4001 frequency points between 17.5 GHz and 26.5 GHz, leading to a frequency sampling interval of 2.25 MHz. Pearson's formula is used for the computation of the correlation, computed between two scalar zero-mean fields $E_1(\bm r_t, \bm \nu )$ and $E_2(\bm r_t, \bm \nu)$ as:
	
	\begin{align}
	C_{1,2}(\bm r_t, \bm r_t) = \frac{\mathbb{E}[(E_1)] \mathbb{E}[(E_2)]}{\sigma_1 \sigma_2}
	\end{align} 
	
	where $\mathbb{E}[E_i]$ and $\sigma_i$ are the expectation and the standard deviation of $E_i$,  respectively, computed over the frequency dimension for each spatial coordinate $\bm r$. The autocorrelation of the first scan is depicted in Fig.~\ref{fig:Xcorr11}.
	
\begin{figure}[ht]
		\centering
		\includegraphics[width=0.9\textwidth]{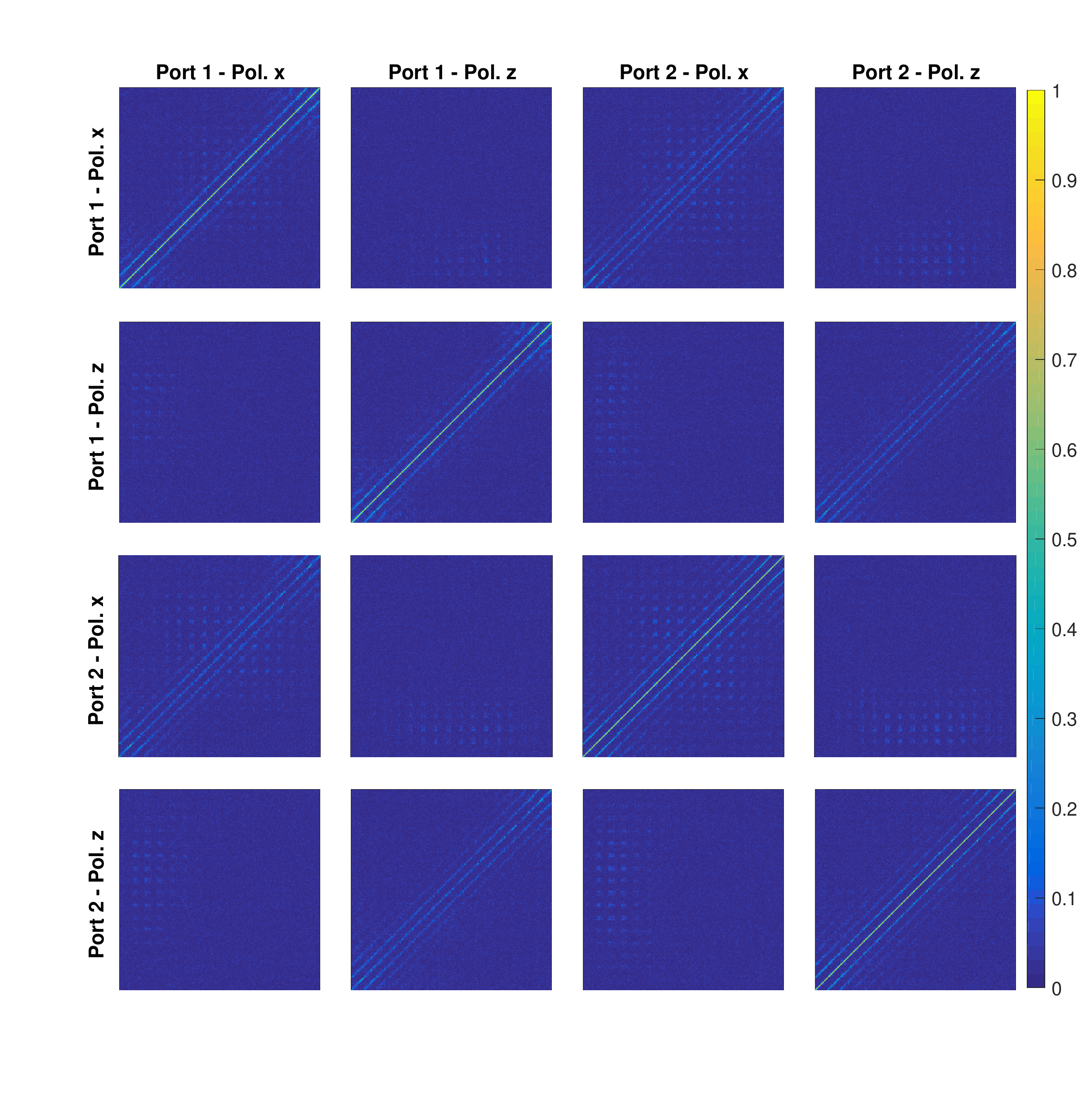}
		\vspace{-1cm}
		\caption{Spatial correlation of the near field scans represented for each couple of ports and polarizations. The axis have been removed to save space and are equivalent for each sub-figure to these of Fig.~\ref{fig:Xcorr11}}
		\label{fig:SpaceCorr}
	\end{figure}

	The pre-processed fields, initially back-propagated to a $18.65 \times 18.65$ cm$^2$ area and sampled using 63 points over the $x$ and $z$ axes, are downsampled to $16 \times 16 $ points per scans for obtaining a spatial grid of $12$ mm, close to the central wavelength of the operating bandwidth $\lambda_{mean} = 13.6$~mm. The two-dimensional coordinate $r_t$ is vectorized on each axis of this figure. The depicted result, corresponding mainly to a diagonal line linking the components sharing the same index, denotes the low level of spatial correlation in this scan obtained by exploiting the frequency diversity in the same way as in a computational imaging experiment. Two low level parallel line are obtained on both sides of the central diagonal, corresponding to the spatial correlation of adjacent points. This phenomenon has already been partially mitigated by resampling the field on a grid closer to the operating wavelength. This study is then extended to both ports and both polarizations, leading to 16 combinations of correlations given in Fig.~\ref{fig:SpaceCorr}.
	
	This global representation of the spatial correlation of the near fields for each couple of ports and polarizations presents the properties expected for interrogating the susceptibility tensor in the target space: the four autocorrelations are close to diagonal matrices, while all the other cross correlations are low. In this way, the polarization information is encoded by the radiation of the metasurface in transmission and reception, and can be retrieved from the inverse formulation by exploiting the frequency diversity.

	\newpage
	
	\section{Practical implementation}
	
\begin{figure}[h]
		\centering
		\includegraphics[width=0.85\textwidth]{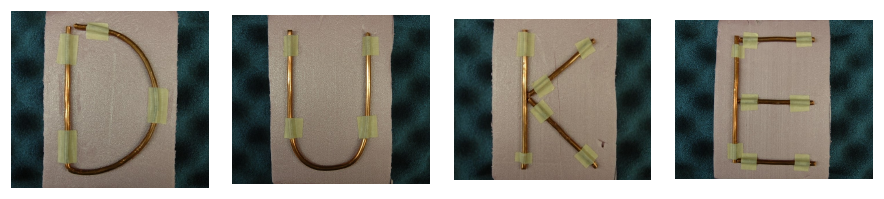}
		\caption{Copper wire letters used as targets. Each letter is $15 \times 9$ $\text{cm}^2$.}
		\label{fig:Wires}
	\end{figure}

	We investigate the polarimetric imaging capabilities of the system using a set of 2 mm thick copper wires as $8\times12\text{ cm}^2$ targets, bent to form the four letters of "DUKE" shown in~Fig.~\ref{fig:Wires}. A target made of thin conductive elements is expected to have an anisotropic scattering according to the orientation of the wires, revealed by the proposed computational imaging system.	The letters are positioned at a distance of 30 cm in front of the radiating aperture and sequential measurements are performed using a vector network analyzer connected to ports 1 and 2 of the cavity. For each measured frequency signal, the inverse problem stated in Eq.~\ref{eq:Inv} is computed using an iterative generalized minimal residual method.
	
	\color{black}
	
\begin{figure}[ht]
		\centering
		\includegraphics[width=\textwidth]{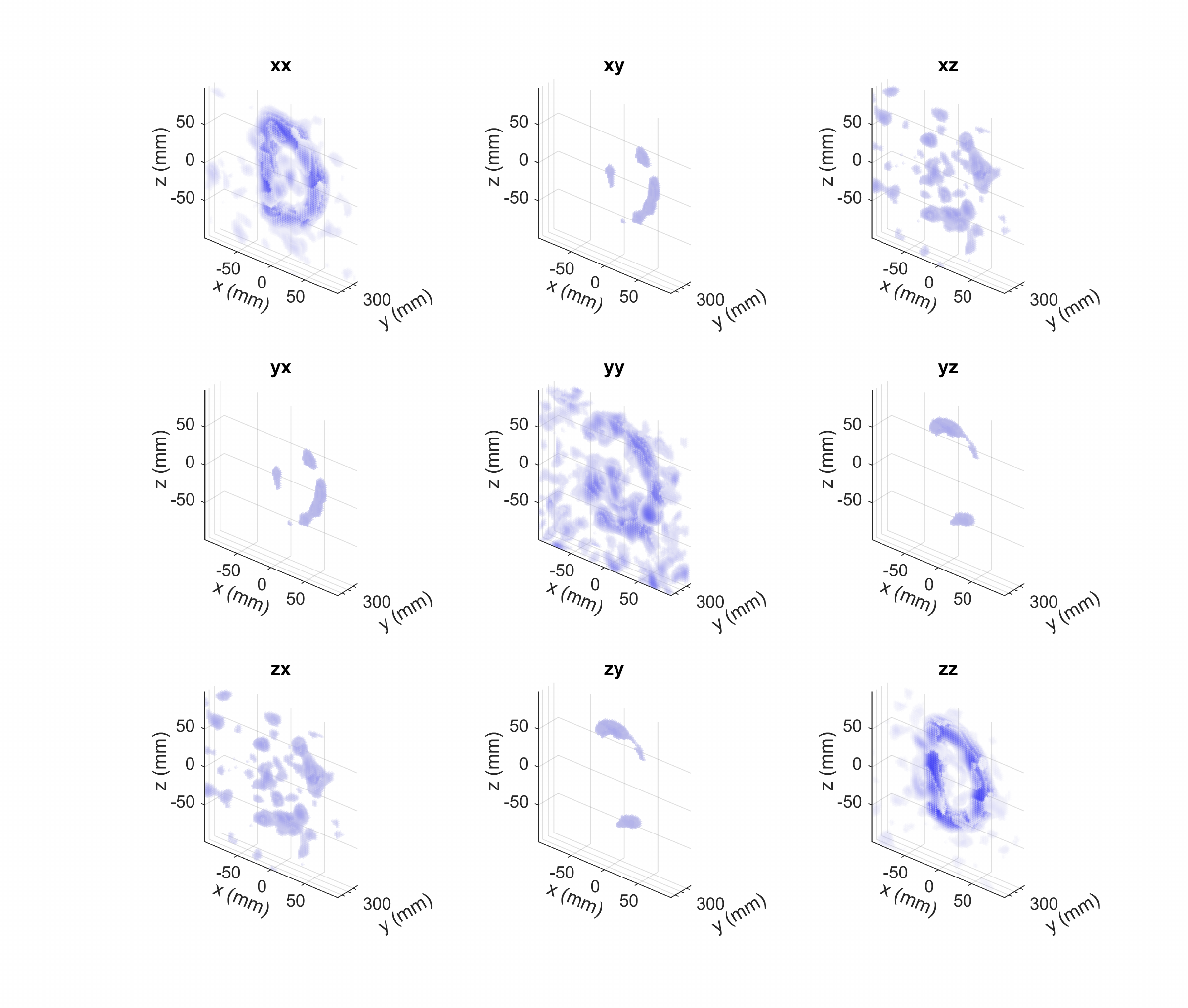}
		\vspace{-0.75cm}
		\caption{Three-dimensional estimation of the full susceptibility tensor $\hat{\bar{\chi}}$ of the first target. The opacity of each voxel is coded on a log scale with a -15~dB minimum threshold.}
		\label{fig:3D_Dtensor}
	\end{figure}

	The estimation of the first susceptibility tensor $\hat{\bar{\chi}}$ is shown in Fig.~\ref{fig:3D_Dtensor} where, first, only the magnitude is displayed. The computation is performed for the $3 \times 3$ interactions between all the polarization components in a target space discretized in $51200$ voxels. The transverse axes $\hat{\bm x}$ and $\hat{\bm z}$ are both sampled on a 2.5 mm grid of 80 elements and the optical axis $\hat{\bm y}$ is sampled on a 5 mm grid of 8 elements. The susceptibility tensor is thus a complex matrix of dimensions $\hat{\bar{\chi}} \in \mathbb{C}^{3\times3\times80\times8\times80}$. The cross-polarized symmetric elements of the tensor are averaged to facilitate numerical processing and to reduce the speckle level.
	
\begin{figure}[ht]
		\centering
		\includegraphics[width=0.97\textwidth]{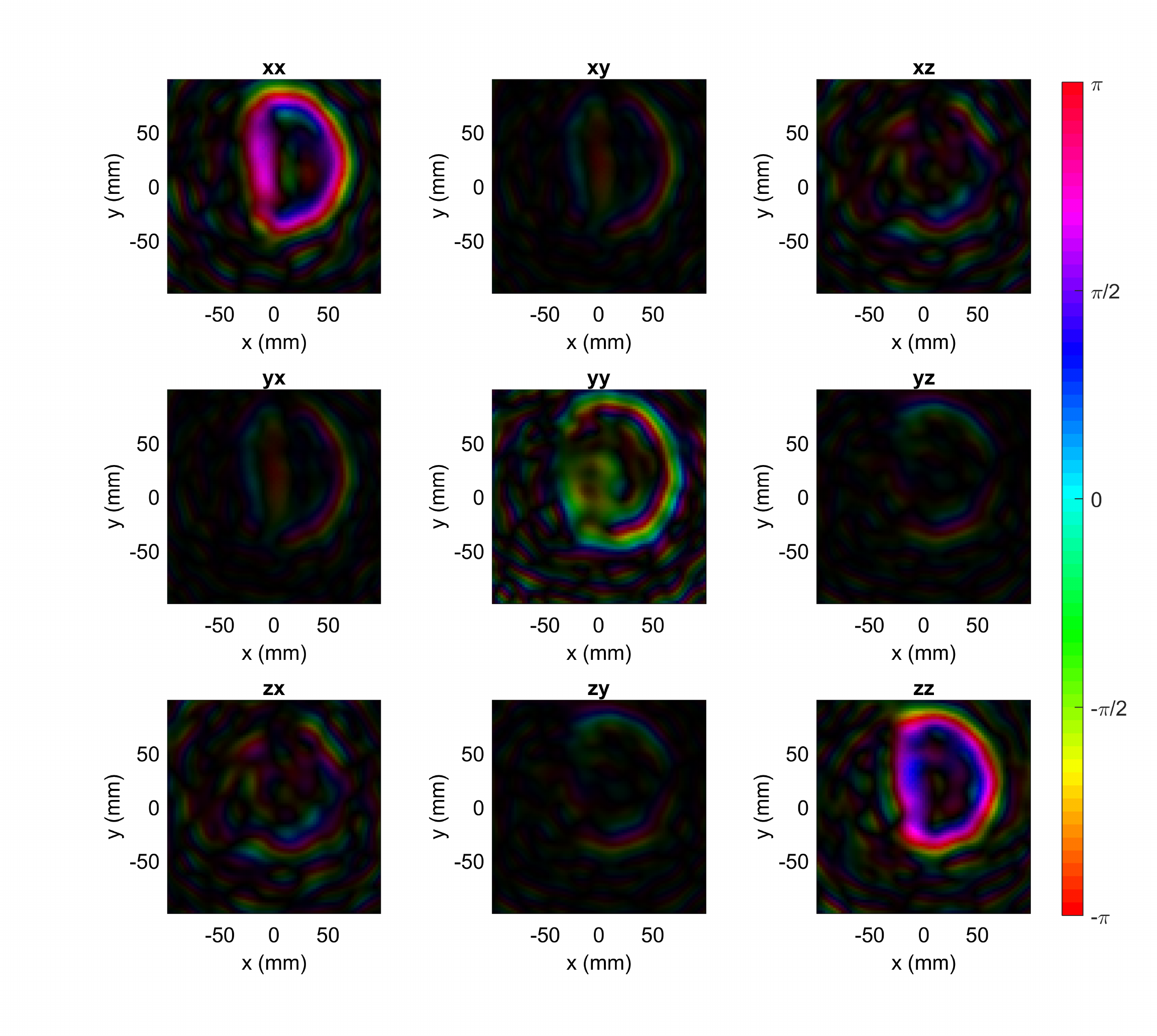}
		\vspace{-0.5cm}
		\caption{Susceptibility tensor extracted from the target plane. The opacity of the pixels corresponds to the linear magnitude of the tensor. The phase is color-coded.}
		\label{fig:2D_Dtensor}
	\end{figure}

	The two transverse co-polarized terms "xx" and "zz" have higher amplitudes and allow for the reconstruction of a continuous target. Interestingly, the contribution of the co-polarized "yy" term is quite significant despite a higher level of speckle. To the best of our knowledge, this feature has never been exploited in conventional polarimetric microwave imaging where most applications are restricted to far-field synthetic aperture radars. In a short-range scenario, the paraxial approximation is no longer valid---leading to an increase amount of measurable information projected along the optical axis. The remaining elements of the susceptibility tensor have a smaller amplitude since they correspond to polarization conversions. This first representation only allows for a study of the magnitude of the tensor in the 3D target space. The resolution of the reconstructed images is directly defined, as in conventional radar imaging, by the size of the radiating aperture and the operating bandwidth~\cite{yurduseven2017frequency}.This analysis is thus continued by extracting the elements in the target plane to consider the impact of both the magnitude and the phase distributions~(Fig.~\ref{fig:2D_Dtensor}). 
	
	\subsection{Correlation of the co-polarized transversed terms}
	
	The results of the two transverse co-polarized images "xx" and "zz" seem similar, but noticeable phase differences are obtained on the horizontal and vertical parts of the target due to the anisotropic nature of the scattering on these thin copper wires. A simple visualization of the anisotropy is thus proposed by computing the correlation $\hat{\chi}_{corr}$ between the two co-polarized transverse terms $\hat{\chi}_{x,x}$ and $\hat{\chi}_{z,z}$ :
	\begin{align}
	\hat{\chi}_{corr}(\bm r) = \hat{\chi}_{x,x}\ \hat{\chi}_{z,z}^*
	\end{align}
	
\begin{figure}[ht]
		\centering
		\includegraphics[width=1\textwidth]{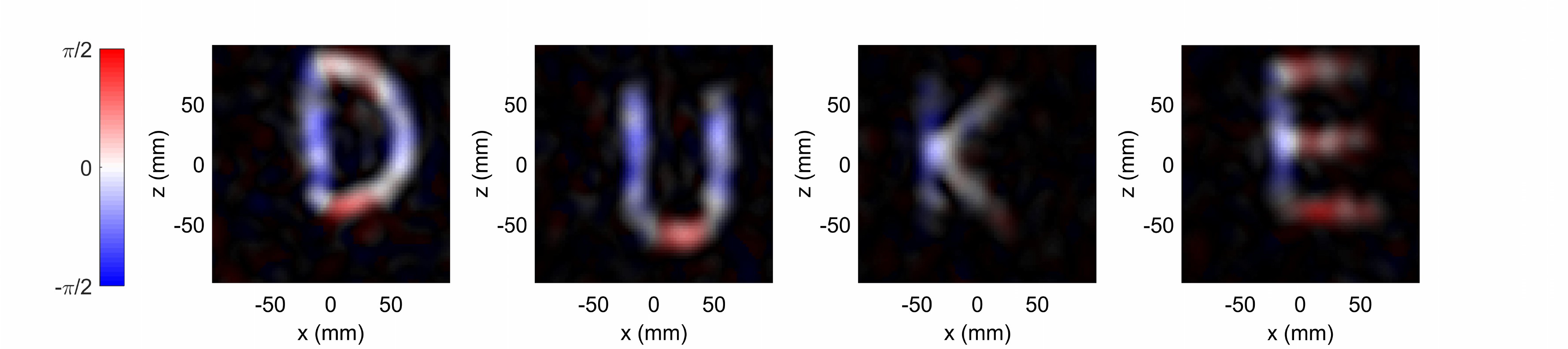}
		\caption{Correlation of two co-polarized transverse components of $\hat{\chi}_{corr}$. The opacity of the figure corresponds to the magnitude of $\hat{\chi}_{corr}$ and the color coding of the phase of $\hat{\chi}_{corr}$.}
		\label{fig:pcolorAnisotropy}
	\end{figure}

	The result is presented in Fig.~\ref{fig:pcolorAnisotropy}, controlling the opacity and the color of each pixel respectively with the magnitude and the phase of $\hat{\chi}_{corr}$. Similar methods are applied in the field of synthetic aperture radar polarimetry by computing covariance matrices for reducing the speckle noise and extracting relevant soil parameters~\cite{lee2009polarimetric,van2011model}.
	
\begin{figure}[ht]
		\centering
		\includegraphics[width=1\textwidth]{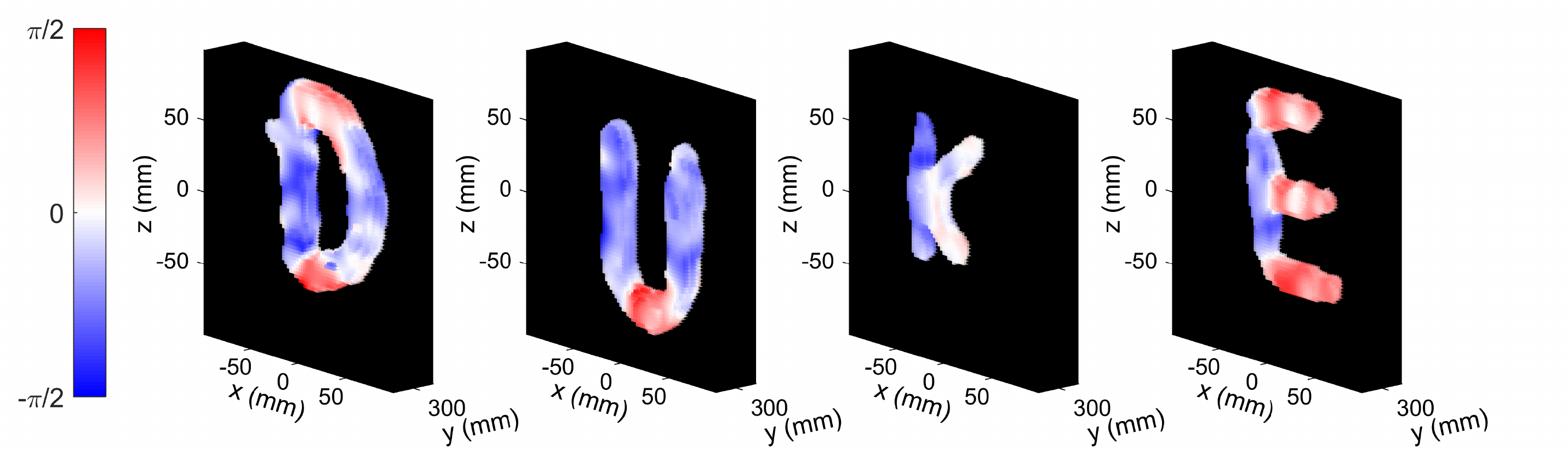}
		\caption{Three-dimensional estimation of $\hat{\chi}_{corr}$. A -20 dB isosurface is represented, color-coded according to the phase of $\hat{\chi}_{corr}$.}
		\label{fig:vol3D_color_coded}
	\end{figure}

	With this simple processing, the orientation of the wires can easily be retrieved according to the phase value, i.e. close to $-\pi/2$ for the vertical direction and $\pi/2$ for the horizontal ones. The elements oriented diagonally are represented by the white color standing for a $0$ radian phase. However, this approach does not allow for the discrimination of the direction of the diagonal orientations of the wires. A comparable analysis can be performed for a three-dimensional visualization of $\hat{\chi}_{corr}$, displaying a -20 dB isosurface of the reconstructed data and using a color coded representation of the phase (Fig.~\ref{fig:vol3D_color_coded}).
	
	Once again, the interference between the two transverse co-polarized terms allows for the reduction of the speckle noise and the visualization of the anisotropic scattering. However, such an approach does not make it possible to take into account all the information contained in the reconstructed susceptibility tensor. A more complex processing is studied based on the eigendecomposition of the susceptibility tensor to directly identify the main axis of each voxel.
	
	\subsection{Eigendecomposition of the susceptibility tensor}
	
\begin{figure}[h!]
		\centering
		\includegraphics[width=1\textwidth]{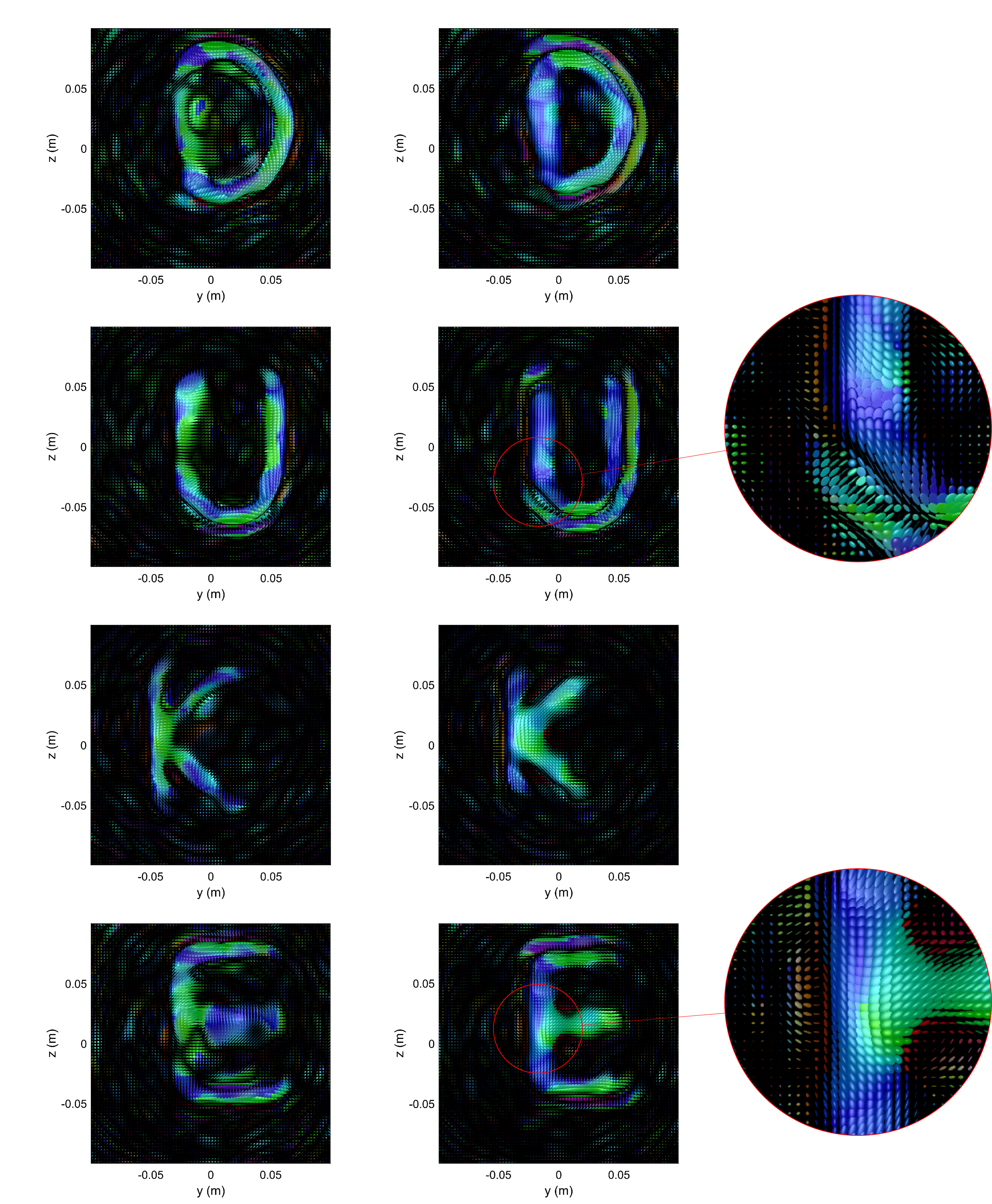}
		\caption{Set of ellipsoids obtained by computing eigendecompositions of the real (left) and imaginary (right) parts of the retrieved susceptibility tensor.}
		\label{fig:Ellipses3D_plan}
	\end{figure}

	The ratio between the small distance of the targets (30 cm) and the size of the radiating aperture (15 cm) makes it possible to obtain polarimetric components oriented along the optical axis. An eigendecomposition is performed on the real and imaginary parts of the retrieved susceptibility tensor. Each voxel of the target space is represented by a 3D ellipsoid oriented and scaled according to the associated eigenvalues and eigenvectors. The ellipsoids are color-coded according to the orientation of the main axis. The results extracted from the target plane are depicted in Fig.~\ref{fig:Ellipses3D_plan}.
	
	In contrast to the previous section where the phase information was used to represent the anisotropic nature of the scattering, this additional analysis relies on the amplitude of the set of components composing the susceptibility tensor, where the real and imaginary parts are processed independently. Some ellipsoids reconstructed on the targets clearly correspond to anisotropic signatures, with an elongated shape corresponding to a eigenvalue larger than the other two, while some voxels have a signature closer to an isotropic response, represented by more spherical ellipsoids. Although the principal axes of the ellipsoids are not necessarily all oriented in the directions of the copper wires, it is possible to discern a direct correlation between the shape of the targets and the colors of the ellipsoids. In particular, the red elements oriented along the optical axis mainly occur for diffraction on the edges of the targets.

	\color{black}
	
	\newpage
	
	\section{Conclusion}
	
	A polarimetric microwave imaging technique has been presented in this paper by extending the computational principle of the scalar approaches that have been previously developed in the literature. The theoretical principle proposed in the introduction suggests a key simplification of the architectures for RF systems that seek to achieve high-resolution images from a single or few source system. The leaky multi-modal cavity used to produce pseudo-orthogonal field patterns in frequency and polarization can be generalized to many metasurface aperture paradigms---all of which are capable of generating field patterns with low correlation as a function of frequency or other parameters.The multiplexing of an increased quantity of information compared to the scalar case requires the development of radiating structures with high quality factors. It is possible in this context to imagine the application of such a technique using a set of metasurfaces operating in a cooperative manner in order to reduce the radiation efficiency and modal diversity constraints imposed on the unique element presented in this proof of concept. In the present example, measurements were taken between two ports of a multi-modal cavity over the band 17.5-26.5 GHz. From the frequency-index measurements, it was possible to reconstruct targets made of copper wires forming the word "DUKE", demonstrating the different modes of operation permitted by the proposed polarimetric approach. Through the two last sections, the anisotropic behavior of the scattering has been highlighted by computing the correlation between the tranverse co-polarized components of the susceptibility tensor, which also allows the reduction of speckle noise, and by computing independent eigendecompositions of the real and imaginary parts of the full susceptibility tensor..
	
	\section*{Funding}
	This work was supported by the Air Force Office of Scientific Research (AFOSR, Grant No. FA9550-12-1-0491).

\end{document}